\documentclass{aa}
\usepackage{graphics}

\begin{document}

   \thesaurus{1(11.09.4; 11.19.3; 13.09.1)}
   \headnote{Letter to the Editor}
   \title{Faint [\ion{O}{iv}] emission from starburst galaxies\thanks{Based on
      observations with ISO, an ESA project with instruments funded
      by ESA Member States and with the participation of
      ISAS and NASA. The SWS is a joint project of SRON and MPE.}}
                                                                     
   \author{D.~Lutz
     \and D.~Kunze
     \and H.W.W.~Spoon
     \and M.D.~Thornley
          }

   \institute{Max-Planck-Institut f\"ur extraterrestrische Physik, Postfach 1603,
             D-85740 Garching, Germany
             }

   \offprints{D.~Lutz\ \ \ (lutz@mpe.mpg.de)}

   \date{Received March 11, 1998; accepted March 27, 1998}

   \maketitle

   \titlerunning{Faint [\ion{O}{iv}] emission from starburst galaxies}
   \authorrunning{D.~Lutz et al.}

   \begin{abstract}

   We report the detection of faint emission in the high-excitation 
   [\ion{O}{iv}] 25.90$\mu$m line in a number
   of starburst galaxies, from observations obtained with the 
   Short Wavelength Spectrometer (SWS) on board ISO. 
   Further observations of \object{M~82} spatially
   resolve the [\ion{O}{iv}] emitting region. 
   Detection of this line in starbursts is 
   surprising since it is not produced in measurable quantities in \ion{H}{ii} 
   regions around hot main-sequence stars, the dominant energy source of
   starburst galaxies. 
   We discuss various models for the formation of this line.
   [\ion{O}{iv}] that is spatially resolved by ISO cannot originate in a 
   weak AGN and must be due to very hot stars or ionizing shocks related to
   the starburst activity. For low-excitation starbursts like \object{M~82}, shocks
   are the most plausible source of [\ion{O}{iv}] emission.


   \end{abstract}

\section{Introduction}

Mid-infrared fine structure lines are powerful probes of dusty and obscured
galactic nuclei, being able to penetrate extinctions up to the equivalent 
of  A$_V\sim 50$. Using the Short Wavelength Spectrometer (SWS) on board the
Infrared Space Observatory (ISO), it is possible to detect faint lines 
and sources. The rich observed spectra can be used for a detailed
modelling of the ionizing spectra of starbursts (e.g. Rigopoulou et al. 
\cite{rigo96}, Kunze et al. \cite{kunze96}) and AGNs (Moorwood et al. 
\cite{moor96}). Clear differences between their spectra make these 
lines a valuable new tool for
discriminating between AGN and starburst activity in visually obscured
galaxies. AGN spectra include emission from highly ionized species and
the so-called coronal lines, requiring photons up to $\sim$300\,eV for their 
creation. 
In contrast, starburst spectra are dominated by lines of low excitation species,
because even hot, massive stars emit few ionizing photons beyond the 
\ion{He}{ii} edge at 54\,eV.

Line ratios like [\ion{O}{iv}]\,25.9$\mu$m / [\ion{Ne}{ii}]\,12.8$\mu$m and 
[\ion{Ne}{v}]\,14.3$\mu$m / [\ion{Ne}{ii}]\,12.8$\mu$m have been used by Lutz et al. 
(\cite{lutz96a}) and Genzel et al. (\cite{genzel98}) to establish the 
dominant source of luminosity in
ultraluminous infrared galaxies (ULIRGs). In
some of the starburst templates studied, 
very faint [\ion{O}{iv}] emission was found, about two orders of magnitude weaker
than in typical AGNs. Faint [\ion{O}{iv}] emission in starbursts is not relevant 
for establishing the power source of ULIRGs, but its origin poses an
interesting problem because its creation ionization energy is slightly
above the \ion{He}{ii} edge.
In this letter we examine possible mechanisms for its production.

\section{Observations and data reduction}

Observations of a variety of starburst galaxies in the [\ion{O}{iv}], [\ion{Ne}{ii}] and
[\ion{Ne}{iii}] 
lines have been obtained with the ISO-SWS in 1996 and 1997, 
as part of a more comprehensive guaranteed time program.
In addition, we use data from a raster of SWS observations along the major 
axis of \object{M~82} 
obtained on March 16, 1996 in an open time program. One of the target lines 
of this program was [\ion{Fe}{ii}] 25.988$\mu$m, which is close enough in
wavelength to extract the 
[\ion{O}{iv}] line from the same scans.  

Integration times of our observations in SWS02 mode were typically 2 seconds
per step, i.e. 200
seconds for the complete up-down scan covering the line. We used standard
procedures from the SWS Interactive Analysis system for data reduction. For
strong sources like bright starbursts, residual fringing is often obvious in the
processed data. This was corrected for by fitting sine functions to line-free
regions of the spectra. This procedure is sufficiently reliable for the
brightest sources, since the 
fringes can be approximated by a single sine function over the small 
observed wavelength range 
near [\ion{O}{iv}], and the lines are narrow. 
Nevertheless, baseline uncertainty due to fringing is often the limiting
factor in measuring the [\ion{O}{iv}] line flux, and allows us to set only
upper limits in some sources. [\ion{Ne}{ii}] and [\ion{Ne}{iii}] 
were always significantly stronger than residual fringes at these wavelengths.

\begin{figure}
\hspace{0.6cm}
\vspace{-0.15cm}
\resizebox{7.5cm}{!}{\includegraphics{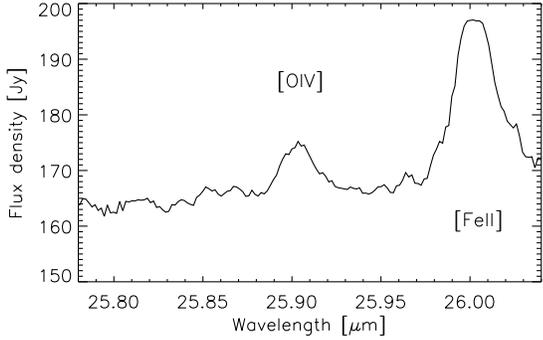}}
\caption{[\ion{O}{iv}] towards the southwest starburst lobe of \object{M~82}}
\label{fig:m82spec}
\end{figure}

\begin{figure}
\hspace{0.6cm}
\vspace{-0.15cm}
\resizebox{7.5cm}{!}{\includegraphics{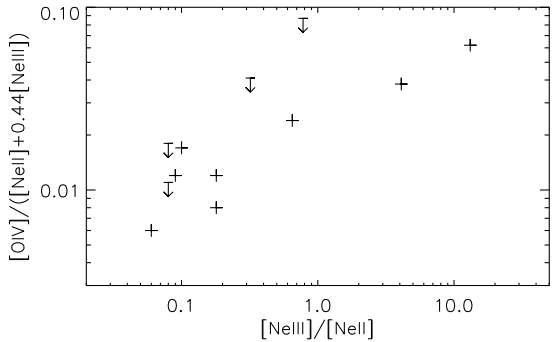}}
\caption{[\ion{O}{iv}]\,25.9\,$\mu$m emission relative to the lower excitation
starburst emission for our sample galaxies, plotted against starburst 
excitation as given by the [\ion{Ne}{iii}]\,15.5\,$\mu$m / 
[\ion{Ne}{ii}]\,12.8\,$\mu$m line ratio}
\label{fig:plotcorr}
\end{figure}

\section{Results}

Fig.~\ref{fig:m82spec}  presents the first detection of [\ion{O}{iv}] 
emission in a starburst, in the spectrum of \object{M~82}. The line fluxes 
measured for all our sources are summarized in Table~\ref{tab:fluxes}, which 
also compares the [\ion{O}{iv}] flux to 
the flux of lower excitation starburst lines. For this comparison, we use 
[\ion{Ne}{ii}]\,12.81$\mu$m + 0.44$\times$[\ion{Ne}{iii}]\,15.55$\mu$m, 
the factor 0.44 chosen
to give equal weight (by mass) to singly and doubly ionized neon. Contrary
to normalizing to just [\ion{Ne}{ii}] or [\ion{Ne}{iii}], this measure will be robust to
changes in excitation of the starburst proper. For two galaxies without 
measured neon lines (\object{NGC 6764} and \object{NGC 6052}), we have estimated the neon fluxes
from the measured [\ion{S}{iii}] 33.48$\mu$m flux using the average scaling for 
the other sample galaxies.

\begin{table}
\caption{Observed line fluxes in starburst galaxies} 
\begin{tabular}{lrr}\hline
Source     &F([\ion{O}{iv}])  &$\rm \frac{[O\,IV]}{[Ne\,II]+0.44\times [Ne\,III]}$\\ 
           &10$^{-20}$\,W\,cm$^{-2}$& \\ \hline
\object{M~82}       &8.00        &0.008           \\
\object{II Zw 40}   &0.55        &0.062           \\
\object{NGC 253}   &5.00        &0.012           \\
\object{IC 342}    &$<$1.00     &$<$0.011        \\
\object{NGC 3256}  &0.93        &0.012           \\
\object{NGC 3690} A&$<$1.20     &$<$0.041        \\
\object{NGC 3690} B/C&0.80      &0.024           \\
\object{NGC 4038}/39$^1$&$<$0.90&$<$0.087        \\
\object{NGC 4945}  &1.40        &0.017           \\ 
\object{M 83}      &0.80        &0.006           \\
\object{NGC 5253}  &0.65        &0.038           \\
\object{NGC 6052}  &$<$0.80     &$<$0.127        \\
\object{NGC 6764}$^2$&$<$0.80     &$<$0.228        \\
\object{NGC 7552}$^2$&$<$1.20     &$<$0.018        \\ \hline       
\end{tabular} \\
$^1$ Interaction region\\
$^2$ [\ion{Ne}{ii}] and [\ion{Ne}{iii}] fluxes estimated from observed 
[\ion{S}{iii}] (see text)
\label{tab:fluxes}
\end{table}

Accurate
linewidths are difficult to determine for the faint [\ion{O}{iv}] lines. All the
detected lines are consistent in width with the starburst fine structure lines
in the  same source, that is unresolved or slightly resolved at the SWS 
resolving power of $\sim$1000.

Faint [\ion{O}{iv}] emission is thus fairly universally detected in starburst
galaxies, at the percent level compared to the starburst neon lines.
There is a weak correlation of [\ion{O}{iv}] strength with excitation of the 
starburst (Fig.~\ref{fig:plotcorr}).
High-excitation, low-metallicity starbursting dwarfs like \object{II Zw 40} and 
\object{NGC 5253} exhibit relatively
stronger [\ion{O}{iv}] than low-excitation starbursts, but the 
effect is not very pronounced compared to the large 
excitation difference as measured by [\ion{Ne}{iii}]/[\ion{Ne}{ii}].

\subsection{Spatially resolved [\ion{O}{iv}] emission in \object{M~82}}

\begin{figure}
\hspace{0.6cm}
\vspace{-0.25cm}
\resizebox{7.5cm}{!}{\includegraphics{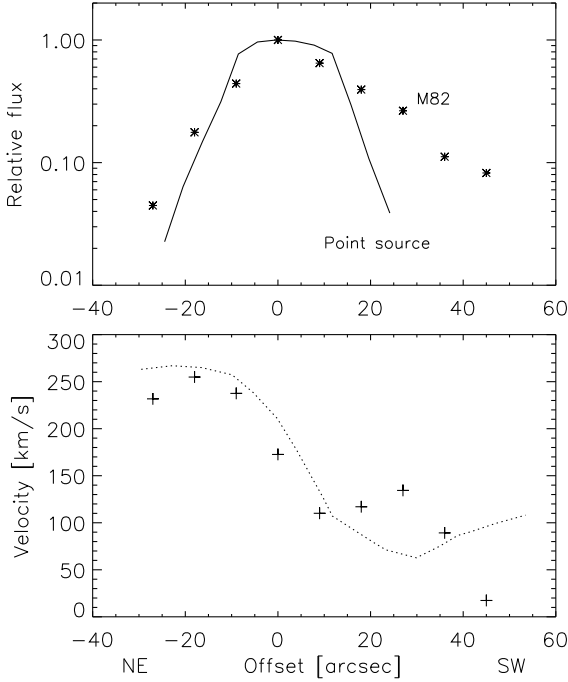}}
\caption{Top: Spatial profile of [\ion{O}{iv}] emission along the major axis of 
\object{M~82} (symbols) compared to the SWS beam profile for a point source (line). 
Bottom: Radial velocity curve for [\ion{O}{iv}] in \object{M~82} (symbols), 
compared to the CO radial velocities as presented by G\"otz et al. (1990).
A shift of -50 km/s has been applied to the CO data.}
\label{fig:m82spatial}
\end{figure}

The SWS02 raster of \object{M~82} was obtained along the major axis
(PA 68\degr ), using a spacing of 9\arcsec. At the time of the observation,
the position angle of the long axis of the SWS apertures was 60.9\degr , i.e.
almost aligned to the major axis. We compared the measured [\ion{O}{iv}] spatial
profile (Fig.~\ref{fig:m82spatial} top) to the SWS beam profile
along the long dimension of the SWS aperture, for the relevant wavelength band
(A. Salama, priv. communication). The [\ion{O}{iv}] emission is clearly 
resolved compared to
the point source beam profile and must originate in a region similar to the 
size of the entire
starburst region. Further support that [\ion{O}{iv}] is resolved comes from the
observation that its `rotation curve' follows the rotation of \object{M~82} 
(Fig.~\ref{fig:m82spatial} bottom).

\section{Discussion}

At a level of just a percent of the strongest low excitation lines, a variety
of possible excitation mechanisms for the [\ion{O}{iv}] line must be considered.

\subsection{Weak AGNs}

Because of the great strength of [\ion{O}{iv}] emission in Seyfert galaxies, quite 
faint and perhaps obscured AGNs embedded in a more luminous starburst would 
contribute sufficient [\ion{O}{iv}], with the additional
constraint that their narrow line width would have to be small. In fact, hard
X-ray observations may indicate AGNs deeply hidden in some of our sources. 
The case is convincing for \object{NGC 4945} (Iwasawa et al. \cite{iwasawa93}), but 
less so for \object{M~82} (Tsuru et al. \cite{tsuru97}). For \object{M~82}, definite 
proof against an AGN origin of 
[\ion{O}{iv}] is provided, however, by the fact that the emitting region is spatially 
resolved and
similar in size to the starburst region. If it were illuminated by a
central source, sufficient [\ion{O}{iv}] could not be produced without exceeding
the observed relatively low [\ion{Ne}{iii}]/[\ion{Ne}{ii}] ratio.

\begin{figure}
\hspace{0.6cm}
\resizebox{7.5cm}{!}{\includegraphics{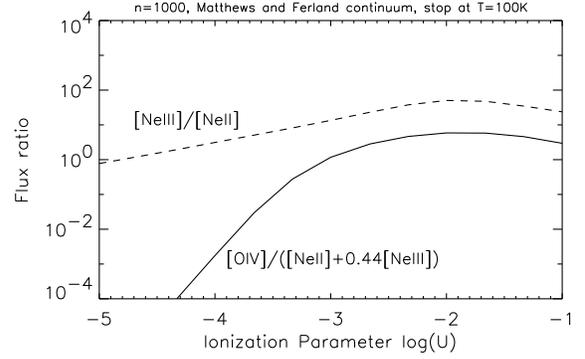}}
\caption{Photoionization models for [\ion{O}{iv}]/([\ion{Ne}{ii}]+0.44\,[\ion{Ne}{iii}]) 
(continuous)
and [\ion{Ne}{iii}]/[\ion{Ne}{ii}] (dashed) as function of ionization parameter in an 
AGN narrow line region} 
\label{fig:agnmodel}
\end{figure}

This constraint is illustrated in Fig.~\ref{fig:agnmodel}, which shows 
predicted line ratios for a simple AGN photoionization model with varying 
ionization parameter computed using
CLOUDY (Ferland \cite{ferland96}). When [\ion{O}{iv}] reaches 1\% of the 
low-excitation
neon lines, [\ion{Ne}{iii}]/[\ion{Ne}{ii}] is already much too high to be consistent
with starbursts like \object{M~82} ([\ion{Ne}{iii}]/[\ion{Ne}{ii}]$\sim$0.17, 
F\"orster-Schreiber et al., in preparation). 
This is a fairly general problem in any
photoionisation scenario, which persists if one adds a small hard 
component to a soft starburst spectrum (as discussed below in the context of 
hot stars). The way to circumvent this problem -- postulate small region(s) 
with very strong [\ion{O}{iv}] but small contribution to the total [\ion{Ne}{ii}] -- 
is not viable here since this would imply a central small 
NLR which is inconsistent with the observations of \object{M~82}. For individual 
galaxies with [\ion{O}{iv}] detections but lacking spatial information, a weak 
central AGN
remains possible. However, we emphasize that the fairly uniform level
of the [\ion{O}{iv}] detections (Fig.~\ref{fig:plotcorr}) requires an 
unlikely finetuning of AGN and starburst activity to fit our sample as a whole.

\subsection{Super-hot stars}

\begin{figure}
\hspace{0.6cm}
\resizebox{7.5cm}{!}{\includegraphics{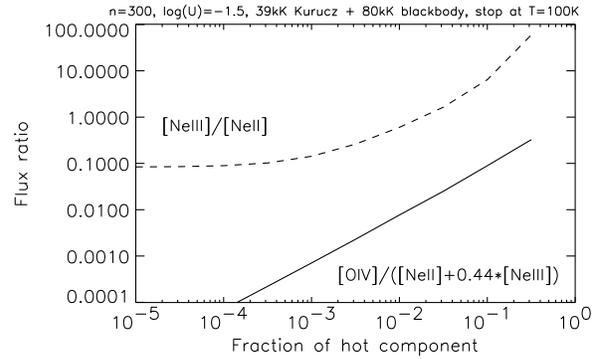}}
\caption{Photoionization models for [\ion{O}{iv}]/([\ion{Ne}{ii}]+0.44\,[\ion{Ne}{iii}]) 
(continuous)
and [\ion{Ne}{iii}]/[\ion{Ne}{ii}] (dashed) in an \ion{H}{ii} region excited by a 39000\,K 
star, with addition of an 80000\,K blackbody contributing different
fractions of the total Lyman continuum luminosity} 
\label{fig:hotstarmodel}
\end{figure}

The ionization edge for creation of [\ion{O}{iv}] is just beyond
the \ion{He}{ii} edge; at higher energies the spectral energy distributions of 
most stars drop
precipitously. However, a small component of hotter (e.g. Wolf-Rayet) stars
might provide the necessary high energy photons.
We have run a photoionization model for an \ion{H}{ii} region excited by
a 39000\,K main sequence star (represented by a Kurucz model atmosphere), 
plus an additional 80000\,K blackbody to represent a harder component.
The blackbody is an ad-hoc choice selected for ease of implementation; however,
other strong sources of photons beyond
54\,eV would give similar results.
As Fig.~\ref{fig:hotstarmodel} shows, this attempt fails to explain [\ion{O}{iv}] 
in low-excitation starbursts since the predicted [\ion{Ne}{iii}]/[\ion{Ne}{ii}] ratio
($\sim$0.5 to 1) exceeds the observations ($\sim$ 0.1) when [\ion{O}{iv}] reaches 
1\% of the low-excitation lines. For the 
high-excitation starbursting dwarfs, such a discrepancy does not arise, and 
hot stars remain an option. 

Again, the inconsistency could be alleviated if {\em small} 
\ion{H}{ii} regions with relatively stronger [\ion{O}{iv}] emission
were dispersed in a lower excitation background. In fact, such
a scenario is qualitatively consistent with the observations, as are
others with distributed local sources of [\ion{O}{iv}].
The major reason to consider it unlikely is that we have failed up 
to now to detect [\ion{O}{iv}] emission even at a {\em similar} level in local star 
forming regions, while we would have to postulate regions with {\em stronger}
emission. The Galactic center, which is closest to starburst galaxies
in many aspects, still shows [\ion{O}{iv}], though even fainter than in the
starbursts (Lutz et al. \cite{lutz96b}). 
In the massive star forming regions \object{W51 IRS2} and \object{30 Doradus}, 
for which [\ion{Ne}{iii}]/[\ion{Ne}{ii}] 
indicates high excitation, we were unable to detect [\ion{O}{iv}] 
at a level of $<$0.01 and
$<$0.005 of [\ion{Ne}{ii}]+0.44[\ion{Ne}{iii}], respectively 
(Thornley et al., in preparation).

\subsection{Planetary nebulae}

High excitation planetary nebulae are a known source of [\ion{O}{iv}] emission.
A young starburst
will, of course, not contain planetary nebulae and it is easy to show that
their integrated contribution from the old stellar population is too faint.
Evolutionary calculations (e.g. Charlot \& Bruzual \cite{charlot91}) 
show that the contribution of post-AGB
stages to the bolometric luminosity  is less than 1\% even in old
populations. Making the extreme assumptions that 10\% of the bolometric
luminosity is due to an old population and that all PAGB objects are like
\object{NGC 7027}, one of the highest excitation planetary nebulae, we estimate a
robust upper limit of 10$^{-20}$\,W\,cm$^{-2}$  for the [\ion{O}{iv}] emission 
from planetary nebulae in \object{M~82}, based on [\ion{O}{iv}] flux, luminosity and distance of 
\object{NGC 7027} as given by Shure et al. (\cite{shure83})
and Beintema et al. (\cite{beintema96}).

\subsection{Ionizing shocks}

There is ample evidence for ionizing shocks in starburst galaxies. Spatially
extended, `Liner'-type optical emission lines can be attributed to shocks, and
kinematic mapping sometimes provides direct evidence for outflowing 
`superwinds' (Heckman et al. \cite{heckman90}). [\ion{O}{iv}] column densities 
approaching 10$^{14}\,$cm$^{-2}$ are
expected for modest velocity shocks (100-200\,km/s, e.g. Shull \& McKee 
\cite{shull79}, 
Dopita \& Sutherland \cite{dopita96}). 
Assuming postshock values of n=1000\,cm$^{-3}$ and T=50000\,K, we estimate a
25.90$\mu$m intensity of
$\sim 3\times 10^{-5}$\,erg\,s$^{-1}$\,cm$^{-2}$\,sr$^{-1}$, equivalent to
$\sim 3\times 10^{-20}$\,W\,cm$^{-2}$ for the SWS beam. For the
assumed conditions, the covering factor of such shocks in the starburst 
region of \object{M~82} would have to be of the order unity. At higher shock 
velocities, the 
[\ion{O}{iv}] column 
would be increasingly dominated by material `at rest' in the 
photoionized precursor in the material ahead of the 
shock front (Dopita \& Sutherland 
\cite{dopita96}). The shock models predict that
intensities similar to those estimated for [\ion{O}{iv}] are emitted 
in optical shock tracers like [\ion{S}{ii}]\,6716/31\AA . This is fully 
consistent
with optical spectroscopy of \object{M~82} (e.g. G\"otz et al. \cite{goetz90}).
We note that the faint shock emission predicted for the
[\ion{Ne}{ii}] and [\ion{Ne}{iii}] lines will be completely dominated by
the emission from \ion{H}{ii} regions.

It is instructive to compare the [\ion{O}{iv}] results for \object{M~82} 
with the SWS observations for \object{RCW 103},
a bright supernova remnant interacting with a dense molecular cloud 
(Oliva et al., in preparation). The [\ion{O}{iv}] intensities are very
similar. The \object{RCW 103} ionic lines are just resolved at the SWS
spectral resolving power, again similar to \object{M~82}. Ionizing shocks hence are a 
plausible origin for the \object{M~82} [\ion{O}{iv}] emission if their total covering
factor approaches unity in the central starburst region of \object{M~82}.

\section{Conclusion}
We have discussed various excitation mechanisms for faint [\ion{O}{iv}] emission 
from starburst galaxies. In general, starburst-related sources and in 
particular ionizing shocks provide the most plausible explanation.
Weak buried AGNs may be plausible for individual sources but 
can be ruled out for the best studied case of \object{M~82} whose [\ion{O}{iv}] emitting
region has been spatially resolved. 
In addition, the fairly small scatter in [\ion{O}{iv}] versus starburst luminosity
favours a starburst-related origin, since no finetuning of two independent 
mechanisms is required.

\begin{acknowledgements}
We thank A.~Sternberg, D.~Rigopoulou and A.~Moorwood for discussions.
SWS and the ISO Spectrometer Data Center at MPE are supported by 
DARA under grants 50 QI 8610 8 and 50 QI 9402 3. 
\end{acknowledgements}

\end{document}